\documentstyle[11pt]{article}
\catcode`\@=11
 
\@addtoreset{equation}{section}
\@addtoreset{equation}{subsection}
\def\theequation{\ifnum\value{subsection}>0\relax
\thesubsection.\arabic{equation}\relax
\else\ifnum\value{section}>0\relax
\thesection.\arabic{equation}\relax
\else\arabic{equation}\fi\fi}

\newcommand{\cR}{{\cal R}}

\begin{document}
\title{\Large \bf Complete Nondiagonal Reflection
Matrices of RSOS/SOS and Hard Hexagon Models}

\author{Changrim Ahn and Choong-Ki You\\
\small Department of Physics\\
\small Ewha Womans University\\
\small Seoul 120-750, Korea}
\date{}

\begin{titlepage}
\maketitle

\begin{abstract}
In this paper we compute the most general nondiagonal reflection matrices 
of the RSOS/SOS models and the hard hexagon model
using the boundary Yang-Baxter equations. 
We find new one-parameter family of the reflection matrices for the RSOS model
in addition to the previous result obtained in \cite{AhnKoo}. 
We also find three classes of the reflection matrices for the SOS model, 
which has one or two free parameters.
For the  hard-hexagon model which can be mapped to RSOS(5) model by folding
four RSOS heights into two,
the solutions can be obtained similarly with a main difference
in the boundary unitarity conditions.
Due to this, the reflection matrices can have two free parameters.
We show that these extra terms can be
identified with the `decorated' solutions.
We also generalize the hard hexagon model by `folding' the RSOS heights
of the general RSOS(p) model and show that they satisfy the integrability
conditions such as the Yang-Baxter and boundary Yang-Baxter equations.
These models can be solved using the results for the RSOS models. 

\end{abstract}
\vspace{2cm}
\rightline{EWHA-TH-006}
\vspace{1cm}
\end{titlepage}

\section{Introduction}
In the study of the two-dimensional integrable models of quantum field 
theories and statistical models, the Yang-Baxter equation (YBE) 
plays essential roles in estabilishing the integrability and
solving the models. 
Recently there has been a lot of efforts in introducing the boundaries
into the integrable systems for possible applications to condensed matter
physics and statistical systems with non-periodic boundary conditions.
The boundaries entail new physical quantity called reflection matrices
which depend on the boundary properties.

The boundary Yang-Baxter equation (BYBE) (also known as the reflection
equation) \cite{cher} is the necessary condition for the integrable statistical
models \cite{skly,oth} and quantum field theories \cite{zam} with boundary.
The equation takes the form
\begin{equation}
R_{1}(u)S_{12}(u^{'}+u)R_2(u^{'})S_{12}(u^{'}-u)=S_{12}(u^{'}-u)R_2(u^{'})
S_{12}(u^{'}+u)R_{1}(u)
\label{eq:bybe1}
\end{equation}
where $R_{1(2)}$ is the boundary reflection matrix in the
auxiliary space $1(2)$ and 
$S_{12}$ is the solution to the YBE.

To date, several solutions of the BYBE have appeared in the literature. 
Compared with the vertex-type models, however, far less is known for the
solution in the face-type model such as
the solid-on-solid (SOS) or restricted-solid-on-solid (RSOS)
model. 
Among these, most of the known solutions are `diagonal' in the sense
that the reflection matrices are diagonal\cite{pea,AhnKoo}.
Using these, one can find the partition functions in the infinite 
lattice limit\cite{miwa}.

Nondiagonal reflection matrices are more interesting because they
can be applied to physical phenomena.
Furthermore, one can find explicit vertex-face correspondence
if one can classify nondiagonal solutions completely.
In addition to those found in \cite{AhnKoo}, 
we will show that different classes of nondiagonal reflection matrices 
are possible. 
These new solutions include one or two free parameters
which are related to those in the boundary potential.
The explicit formulae between two sets of parameters are still not clear.
We hope our complete nondiagonal solutions for the SOS model may 
be related explicitly with those of the boundary sine-Gordon model.

>From a physical point of view, many interesting models are face-type;
the RSOS/SOS models, the hard-hexagon model (HHM) etc.
These models play very important roles in the
statistical mechanics system and in the quantum field theories
such as the perturbed conformal field theories.

In this paper we derive complete nondiagonal reflection matrices
for the RSOS/SOS and the HHM in a unified way. 
We will express BYBE as a linear equation which should satisfy
extra nontrivial conditions.
Due to the linearity, the general solutions are 
linear combinations of each solution with arbitrary coefficients.
Some of these coefficients are fixed
by the boundary crossing and unitarity conditions \cite{zam}.
The reflection matrices of the HHM can be constructed from those
of the RSOS model since the HHM can be mapped to 
the RSOS(5). We consider a similar mapping for the generic RSOS(p) model
and their boundary reflection matrices.

\section{RSOS(p) model}

In this section we solve the BYBE for the 
$\mbox{RSOS}(p)\;;$ $p=3,4\ldots$ scattering theory. The $\mbox{RSOS}(p)$ 
scattering theory is based on a $(p-1)$~-~fold degenerate vacuum structure, 
vacua can be associated with nodes of the ${\cal A}_{p-1}$ Dynkin diagram.
The quasi particles  in the scattering theory are kinks that interpolate 
neighboring vacua, they can be denoted by non-commutative symbols 
$K_{ab}(u)$ where $|a-b|=1$ with $a,b=1,\ldots,p-1$ and 
$u$ is related to the the kink 
rapidity $\theta$ by $u=-i\theta/p$, so that the physical strip is given by 
$0<{\rm Re}\;u<\pi/p$. In the rest of the paper, we will refer to $a,b$
as heights or spins.
Formally, scattering between two kinks can be represented by the following 
equation
\begin{equation}
K_{da}(u)K_{ab}(u^{'})=\sum_{c}S^{ab}_{dc}(u-u^{'})
K_{dc}(u^{'})K_{cb}(u)\;,
\end{equation}
where the $S$-matrix is given by
\begin{equation}
S^{ab}_{dc}(u)={\cal U}(u)
\left(\frac{[a][c]}{[d][b]}\right)^{-u/2\gamma}W^{ab}_{dc}(u) \;.
\label{eq:bulk}
\end{equation}
The Boltzmann weight,
\begin{equation}
W^{ab}_{dc}(u)=
\sin u \delta_{bd}\left(\frac{[a][c]}{[d][b]}\right)^{1/2}+
\sin(\gamma-u)\delta_{ac}\;,
\end{equation}
satisfies the YBE in the RSOS representation.

Here $[a]$ denotes the usual $q$-number,
\[
[a]=\frac{\sin(a\gamma)}{\sin\gamma}\qquad
\gamma=\frac{\pi}{p} \;,
\]
and the overall factor ${\cal U}(u)$ is a product of Gamma functions,
\begin{eqnarray}
{\cal U}(u)&=&\frac{1}{\pi}\Gamma\left(\frac{\gamma}{\pi}\right)
\Gamma\left(1-{u\over{\pi}}\right)
\Gamma\left(1-{\gamma\over{\pi}}+{u\over{\pi}}\right)
\prod^{\infty}_{l=1}{F_l(u)F_l(\gamma-u)\over{F_l(0)F_l(\gamma)}},
\nonumber\\
F_l(u)&=&{\Gamma\left({2l\gamma\over{\pi}}-{u\over{\pi}}\right)
\Gamma\left(1+{2l\gamma\over{\pi}}-{u\over{\pi}}\right)\over{
\Gamma\left({(2l+1)\gamma\over{\pi}}-{u\over{\pi}}\right)
\Gamma\left(1+{(2l-1)\gamma\over{\pi}}-{u\over{\pi}}\right)}}\;.
\label{eq:uf}
\end{eqnarray}
This factor, satisfying the relations
\begin{eqnarray*}
{\cal U}(u){\cal U}(-u)\sin(\gamma-u) \sin(\gamma+u)&=&1\\
{\cal U}(\gamma-u)&=&{\cal U}(u)\;,
\end{eqnarray*}
together with the overall $q$-number factor ensures
that the $S$-matrix satisfies both crossing and unitarity
constraints:
\begin{eqnarray}
S^{bc}_{ad}(\gamma-u)&=&S^{ab}_{dc}(u)
\\
\sum_{c^{'}}S^{ab}_{dc^{'}}(u)S^{c^{'}b}_{dc}(-u)&=&\delta_{ac}\;.
\end{eqnarray}

Let us consider now the above scattering theory
in the presence of a boundary.
The scattering between the kink and
the boundary denoted formally by ${\bf B}_{a}$
is described by the equation
\begin{equation}
K_{ab}(u){\bf B}_{a}=\sum_c R^b_{ac}(u) K_{bc}(-u){\bf B}_{c} \;.
\end{equation}
Note that in this representation,
the boundary naturally carries an RSOS spin.

The function $R^{b}_{ac}$ is called the boundary
reflection matrix and satisfies the BYBE,
which in the RSOS representation takes the form
\begin{eqnarray}
\lefteqn{\sum_{a^{'},b^{'}}R^a_{bb^{'}}(u)S^{ac}_{b^{'}a^{'}}(u^{'}+u)
R^{a'}_{b^{'}b^{''}}(u^{'})S^{a^{'}c}_{b^{''}a^{''}}(u^{'}-u)=}
\nonumber\\
&&
\sum_{a^{'},b^{'}}S^{ac}_{ba^{'}}(u^{'}-u)
R^{a^{'}}_{bb^{'}}(u^{'})S^{a^{'}c}_{b^{'}a^{''}}(u^{'}+u)
R^{a''}_{b^{'}b^{''}}(u)\;. \label{eq:bybe}
\end{eqnarray}
In general, the function $R^{a}_{bc}(u)$ can be written as
\begin{equation}
R^{a}_{bc}(u)={\cal R}(u)\left(\frac{[b][c]}{[a][a]}\right)^{-u/2\gamma}
\left[\delta_{b\neq c}X^{a}_{bc}(u)+\delta_{bc}
\left\{\delta_{b,a+1}U_a(u)+\delta_{b,a-1}D_a(u)\right\}\right]\;,
\label{eq:boundary}
\end{equation}
where ${\cal R}(u)$ has to be determined from the boundary crossing and
unitarity constraints, while $X^{a}_{bc}$ and $U_a,D_a$ have to be 
determined from the BYBE.  An overall $q$-number factor has also been 
multiplied to the above to cancel that from the bulk $S$-matrix 
in order to simplify the BYBE. 
If $X^{a}_{bc}$ does not vanish, the boundary $R$-matrix describes 
non-diagonal
scattering process, otherwise the scattering is called diagonal.
Note that due to the restriction
that vacuum assumes value $1,\ldots,p-1$, 
$X^{1}_{bc},X^{p-1}_{bc},D_1,U_{p-1}$
are not defined. The case $p=3$
has only diagonal scattering, so $X^a_{bc}$ does not exist.

We will concentrate in what follows to the scattering where the off-diagonal 
component $X^a_{bc}$ is non-vanishing. To start, 
the case $b\neq c\neq b^{''}$ in
eqn.(\ref{eq:bybe}) gives
\begin{equation}
X^a_{a-1,a+1}(u^{'})X^{a+2}_{a+1,a+3}(u)=X^a_{a-1,a+1}(u)
X^{a+2}_{a+1,a+3}(u^{'})\;;2\leq a\leq p-4 \;.
\end{equation}
This equation implies that $X^{a}_{a\pm 1,a\mp 1}$ can be written as
\[
X^{a}_{a\pm 1,a\mp 1}(u)=h_{\pm}(u)X_{\pm}^{a} \;,
\]
where $h_{\pm}(u)$ depends only on $u$ and 
$X^{a}_{\pm}$ only on $a$.

On the other hand, the case $c=b=b^{''}, a=a^{''}$ gives
\begin{equation}
X^a_{a-1,a+1}(u^{'})X^{a}_{a+1,a-1}(u)=X^a_{a-1,a+1}(u)
X^{a}_{a+1,a-1}(u^{'})\;;2\leq a\leq p-2 \;,
\end{equation}
which implies that 
\[
h_{+}(u^{'})h_{-}(u)=h_{+}(u)h_{-}(u^{'})\;,
\]
from which we conclude that 
\[
h_{+}(u)={\rm (const.)}h_{-}(u)\;.
\]
Absorbing the constant in the above equation into $X^{a}_{-}$ or $X^a_{+}$, 
we can make $h_{+}$ equal to $h_{-}$ so that we can absorb 
the $h_{\pm}(u)$ into the overall $\cR(u)$ factor and
treat $X^{a}_{bc}$ as $u$ independent from now on.

With this simplification, eqn.(\ref{eq:bybe}) can be broken
down into the following independent equations in addition to the above
two equations:
\begin{eqnarray}
&&U^{'}_{a}D_{a+2}f_{+}\left(1+f_{-}{[a]\over{[a+1]}}\right)+
D^{'}_{a+2}D_{a+2}f_{-}\left(1+f_{+}{[a+2]\over{[a+1]}}\right)\nonumber \\
&&\mbox{}+X^{a+2}_{a+1,a+3}X^{a+2}_{a+3,a+1}f_{-}=U_{a}D^{'}_{a+2}f_{+}
\left(1+f_{-}{[a+2]\over{[a+1]}}\right) \nonumber \\
&&\mbox{}+U^{'}_{a}U_{a}f_{-}\left(1+f_{+}{[a]\over{[a+1]}}\right)
+X^{a}_{a-1,a+1}X^{a}_{a+1,a-1}f_{-}\label{eq:s0}
\end{eqnarray}
for $1\leq a\leq p-3$, 
\begin{eqnarray}
&&D^{'}_{a+1}f_{-}\left(1+f_{+}{[a+1]\over{[a]}}\right)+
U^{'}_{a-1}f_{+}\left(1+f_{-}{[a-1]\over{[a]}}\right)\nonumber\\
&&\mbox{\hspace{1cm}}=U_{a-1}f_{+}-U_{a+1}f_{-}\label{eq:s1a}\\
&&\nonumber\\
&&U^{'}_{a}f_{-}\left(1+f_{+}{[a]\over{[a+1]}}\right)+
D^{'}_{a+2}f_{+}\left(1+f_{-}{[a+2]\over{[a+1]}}\right)\nonumber\\
&&\mbox{\hspace{1cm}}=D_{a+2}f_{+}-D_{a}f_{-}\label{eq:s1b}
\end{eqnarray}
for $2\leq a\leq p-3$, and 
\begin{eqnarray}
&&U^{'}_{a-2}f_{+}f_{-}{[a][a-2]\over{[a-1]^2}}-U^{'}_{a}
+D^{'}_{a}\left(1+f_{-}{[a]\over{[a-1]}}\right)
\left(1+f_{+}{[a]\over{[a-1]}}\right)\nonumber\\
&&\mbox{\hspace{1cm}}=D_{a}\left(1+f_{+}{[a]\over{[a-1]}}\right)
-U_{a}\left(1+f_{-}{[a]\over{[a-1]}}\right) 
\label{eq:s2a}\\&&\nonumber \\
&&D^{'}_{a+2}f_{+}f_{-}{[a][a+2]\over{[a+1]^2}}-D^{'}_{a}+
U^{'}_{a}\left(1+f_{-}{[a]\over{[a+1]}}\right)
\left(1+f_{+}{[a]\over{[a+1]}}\right)\nonumber\\
&&\mbox{\hspace{1cm}}=U_{a}\left(1+f_{+}{[a]\over{[a+1]}}\right)
-D_{a}\left(1+f_{-}{[a]\over{[a+1]}}\right)\label{eq:s2b}
\end{eqnarray}
for $2\leq a\leq p-2$. In the above equations, we used a compact notation
where
$U_a=U_a(u),\ U^{'}_a=U_a(u^{'})$ (similarly for $D_a$)
and 
\[f_{\pm}=\sin(u^{'}\pm u)/\sin(\gamma-u^{'}\mp u)\;.\]

In addition, it should also be mentioned that the last term 
in the rhs (lhs) of eqn.(\ref{eq:s0}) is present only when $a\neq 1 (p-3)$
and the first terms of eqns.(\ref{eq:s2a}) and (\ref{eq:s2b}) are 
allowed only for $a\neq 2$ and $a\neq p-2$, respectively. 

Before solving the equations directly,
it is helpful to investgate the structure of BYBE.
First of all, the BYBE is covariant under the transformation
\begin{equation}
a \rightarrow p-a \;,U_a(u) \rightarrow \pm D_{p-a}(u)\;,
\label{eq:cov}
\end{equation}
and have the following symmetry
\begin{equation}
U_a(u)=-D_a(-u)\label{eq:sym}
\end{equation}
for $2\leq a\leq p-2$.
Futhermore, since the eqns.(\ref{eq:s1a})-(\ref{eq:s2b}) are linear,
their general solutions are linear combinations of `fundamental' ones.
Another important fact is that the amplitudes with $a$ even and those with
$a$ odd are completely decoupled in BYBE and give different solutions 
in general.
However if $p$ is odd, two sets are related by (\ref{eq:cov}) and 
have the same solutions.

By solving the linear equations (\ref{eq:s1a})-(\ref{eq:s2b}), 
we find the most general nondiagonal solution for $p \geq 5$:
\begin{eqnarray}
U_a(u)&=&A\sin(2u+a\gamma)+\frac{B}{\sin 2u}
+\frac{\epsilon_p\epsilon_{a-1}C}{\sin a\gamma}
\nonumber
\\
&&+\frac{\epsilon_{p-1}D}{\sin a\gamma}
\left\{ \frac{\sin(2u+a\gamma)}{\sin 2u} -(-1)^a\right\}
\nonumber\\
D_{a}(u)&=&A\sin(2u-a\gamma)+\frac{B}{\sin 2u}
-\frac{\epsilon_p\epsilon_{a-1}C}{\sin a\gamma}
\label{eq:udn}\\
&&-\frac{\epsilon_{p-1}D}{\sin a\gamma}
\left\{ \frac{\sin(2u-a\gamma)}{\sin 2u} -(-1)^a\right\} \;,
\nonumber
\end{eqnarray}
where $\epsilon_a$ is $0(1)$ if $a$ is odd(even).
and $A,B,C$ and $D$ are free parameters.

Having found $U_a,D_a$, the function $X^a_{bc}$ can be 
easily obtained from eqn.(\ref{eq:s0}),
after taking $u^{'}$ to be $-u$ since $X^a_{bc}$
does not depend on the rapidity. This gives
\[X^{a+2}_{a+1,a+3}X^{a+2}_{a+3,a+1}-X^{a}_{a-1,a+1}X^{a}_{a+1,a-1}
=U_a(-u)U_a(u)-D_{a+2}(-u)D_{a+2}(u)
\]
for $1 \leq a \leq p-3$.
Substituting $U_a,D_a$ into the rhs and iterating the equations, we get
\begin{eqnarray}
X^{a}_{a-1,a+1}X^{a}_{a+1,a-1}&=&
\epsilon_p\epsilon_aX^2_{13}X^2_{31}
+A^2\left\{ \sin^2 (1+\epsilon_p\epsilon_a)\gamma -\sin^2  a\gamma\right\}
\nonumber
\\
&&-2AB\left\{\cos (1+\epsilon_p\epsilon_a)\gamma -\cos a\gamma \right\}
\nonumber
\\
&&+\epsilon_p\epsilon_{a-1} C^2 \left\{ \frac{1}{\sin^2 \gamma}
-\frac{1}{\sin^2 a\gamma} \right\}
\label{eq:xn}
\\
&&+2\epsilon_{p-1} D^2
\left\{ \frac{1}{\sin^2 \gamma} -\frac{1}{\sin^2 a\gamma} \right\}
\nonumber
\\
&&+2\epsilon_{p-1} D^2
\left\{ \frac{\cos\gamma}{\sin^2 \gamma}
+(-1)^a \frac{\cos a\gamma}{\sin^2 a\gamma} \right\}\;.
\nonumber
\end{eqnarray}
Since this equation fixes only the product, $X^{a}_{a-1,a+1}$
and $X^{a}_{a+1,a-1}$ are determined upto a gauge factor.
Inserting $a=1$ one gets $X^1_{02}X^1_{20}=0$ as expected.
For $p$ even, $X^2_{13}X^2_{31}$ is not determined yet.

Consider now boundary unitarity and crossing symmetry conditions 
for the reflection matrix $R^a_{bc}(u)$.
Due to these conditions, the overall factor $R(u)$ 
should satisfy
\begin{eqnarray}
\sum_{c}R^{a}_{bc}(u)R^{a}_{cd}(-u)&=&\delta_{bd}\\
\sum_{d}S^{ac}_{bd}(2u)R^{d}_{bc}(\gamma/2+u)&=&R^{a}_{bc}(\gamma/2-u)\;.
\end{eqnarray}

In terms of (\ref{eq:boundary}) and (\ref{eq:sym}), the unitarity
condition becomes
\begin{eqnarray*}
&&{\cal R}(u){\cal R}(-u)\left[X^a_{a+1,a-1}X^a_{a-1,a+1}
-U_a(u)D_a(u)\right]=1 \;;2\leq a \leq p-2
\\
&&\cR(u)\cR(-u)U_1(u)U_1(-u)=1
\\
&&\cR(u)\cR(-u)D_{p-1}(u)D_{p-1}(-u)=1 \;.
\end{eqnarray*}
By inserting eqns.(\ref{eq:udn}) and 
(\ref{eq:xn}) to the above, we find the following nontrivial constraints
for the free parameters:
\begin{eqnarray*}
&i)& AB=AC=AD=0 \\
&ii)& X^2_{13}X^2_{31}=
A^2(\sin^2 \gamma -\sin^2 2\gamma)
+\frac{C^2}{\sin^2 \gamma}\qquad{\rm for\ even}\ p \;.
\end{eqnarray*}
Note that all the cases satisfying the constraints,
have at most one free parameter since we can absorb overall constant
into ${\cal R}(u)$.
We list each class of the reflection matrices as follows:
\\
Class (I) ~~~for general $p\ (p\neq 3,4)$: $B=C=D=0,\; A=1$
\begin{eqnarray}
U_a(u) &=& \sin (2u+a\gamma)
\nonumber
\\
D_a(u) &=& \sin (2u-a\gamma)
\\
X^a_{a-1,a+1}X^a_{a+1,a-1}&=& \sin^2 \gamma-\sin^2 a\gamma \;.
\nonumber
\end{eqnarray}

These weights have $U_{p-a}(u)=-D_a(u)$ symmetry and no free parameter.
This solution is the one obtained in \cite{AhnKoo}.
The unitarity condition gives
\begin{equation}
\cR(u)\cR(-u) (-\sin^2 2u +\sin^2 \gamma) =1 \;.
\label{eq:r1}
\end{equation}
The crossing symmetry condition becomes
\begin{equation}
{\cal U}(2u){\cal R}(\gamma/2+u)\sin(\gamma-2u)=
{\cal R}(\gamma/2-u) \;.\label{eq:r2}
\end{equation}

The factor ${\cal R}(u)$ can be determined from eqns.(\ref{eq:r1}),
(\ref{eq:r2}) up to the usual CDD ambiguity by separating  
${\cal R}(u)={\cal R}_0(u){\cal R}_1(u)$ where ${\cal R}_0$ satisfies
\begin{eqnarray}
\cR_{0}(u)\cR_0(-u)&=&1
\nonumber
\\
{\cal U}(2u)\cR_0(\gamma/2+u)\sin(\gamma-2u)&=&\cR_0(\gamma/2-u)\;,
\end{eqnarray}
whose minimal solution reads
\[
\cR_0(u)=\frac{F_0(u)}{F_0(-u)} \;.
\]
While $\cR_1$ satisfies
\begin{eqnarray}
&&\cR_1(u)\cR_1(-u)(-\sin^2 2u +\sin^2 \gamma)=1
\nonumber
\\
&&\cR_1(u)=\cR_1(\gamma-u)
\end{eqnarray}
with minimal solution
\[
\cR_1(u)=\frac{1}{2}\sigma(\gamma/2,u)\sigma(\pi/2-\gamma/2,u)
\;.
\]

Here $\sigma(x,u)$ is a well-known building block satisfying the relations
\begin{eqnarray*}
\sigma(x,u)&=&\sigma(x,\gamma-u)
\\
\sigma(x,u)\sigma(x,-u)&=&\left[ \cos(x+u) \cos(x-u) \right]^{-1}\;,
\end{eqnarray*}
and is given by
\begin{eqnarray*}
\sigma(x,u)&=&\frac{\textstyle\prod(x,\frac{\gamma}{2}-u)
\prod(-x,\frac{\gamma}{2}-u)
\prod(x,-\frac{\gamma}{2}+u)\prod(-x,-\frac{\gamma}{2}+u)}
{\textstyle\prod^2(x,\frac{\gamma}{2})\prod^2(-x,\frac{\gamma}{2})}\\&&\\
\prod(x,u)&=&\displaystyle\prod_{l=0}^{\infty}
\frac{\textstyle\Gamma\left(\frac{1}{2}+(2l+\frac{1}{2})\frac{\gamma}{\pi}
+\frac{x}{\pi}-\frac{u}{\pi}\right)}
{\textstyle\Gamma\left(\frac{1}{2}+(2l+\frac{3}{2})
\frac{\gamma}{\pi}+\frac{x}{\pi}-\frac{u}{\pi}\right)}\;.
\end{eqnarray*}
\\
\\
Class (II) ~~~for even $p\ (p\neq 4)$: $A=0,\;C=1$
\begin{eqnarray}
U_a(u) & = & \frac{B}{\sin 2u} +\frac{\epsilon_{a-1}}{\sin a\gamma}
\nonumber
\\
D_a(u) & = & \frac{B}{\sin 2u} -\frac{\epsilon_{a-1}}{\sin a\gamma}
\\
X^a_{a-1,a+1}X^a_{a+1,a-1}
 & = & \frac{1}{\sin^2 \gamma} -\frac{\epsilon_{a-1}}{\sin^2 a\gamma}\;.
\nonumber
\end{eqnarray}

This solution satisfies $U_{p-a}(u)=U_p(u)$
(similarly for $D_a(u)$ and $X^a_{bc}$) and include one free parameter.
To fix the overall factor $\cR(u)$, $\cR_0(u)$ is the same as Class (I)
while the $\cR_1(u)$ satisfies
\begin{equation}
\cR_1(u)\cR_1(-u)
\left(\frac{1}{\sin^2 \gamma} -\frac{B^2}{\sin^2 2u} \right)=1.
\end{equation}
The minimal solution is 
\[
\cR_1(u)=\sin \gamma\;
\frac{\sigma(x,u)\sigma(\pi/2-x,u)}{\sigma(0,u)\sigma(\pi/2,u)} \;,
\]
where
\[
\sin 2x =B\sin\gamma \;.
\]
\\
Class (III) ~~~for odd $p\ (p\neq 3)$: $A=0,\;D=1$
\begin{eqnarray}
U_a&=& \frac{B}{\sin 2u} +
\frac{1}{\sin a\gamma}\left\{\frac{\sin(2u+a\gamma)}
{\sin 2u} -(-1)^a \right\}
\nonumber
\\
D_a&=&\frac{B}{\sin 2u} -\frac{1}{\sin a\gamma}
\left\{ \frac{\sin(2u-a\gamma)}{\sin 2u} -(-1)^a \right\}
\\
X^a_{a-1,a+1}X^a_{a+1,a-1}&=&
2\left\{\frac{1}{\sin^2 \gamma}-\frac{1}{\sin^2 a\gamma}\right\}
+2\left\{\frac{\cos\gamma}{\sin^2 \gamma}
+(-1)^a \frac{\cos a\gamma}{\sin^2 a\gamma} \right\} \;.
\nonumber
\end{eqnarray}

This solution has $U_{p-a}(u)=D_p(u)$ symmetry and one free parameter.
While $\cR_0(u)$ does not change, the $\cR_1(u)$ satisfies
\begin{eqnarray*}
\cR_1(u)\cR_1(-u)
\left(\frac{1}{\sin^2 \frac{\gamma}{2}}
-\frac{2B \cos 2u}{\sin^2 2u}
-\frac{1+B^2}{\sin^2 2u}
\right) =1.
\end{eqnarray*}
The minimal solution is
\[
\cR_1(u)=\sin \frac{\gamma}{2}\;\;
\frac{\sigma(x_1,u)\sigma(x_2,u)}{\sigma(0,u)\sigma(\pi/2,u)}
\;, \]
where
\begin{eqnarray*}
\cos^2 x_1 +\cos^2 x_2 &=&1+B\sin^2 \frac{\gamma}{2}
\\
\cos x_1 \cos x_2 &=& \frac{1}{2}(1+B)\sin \frac{\gamma}{2} \;.
\end{eqnarray*}

In the above analysis, we omit special cases of $p=3,4$
since $p=3$ has only the diagonal reflection matrices and $p=4$
has been extensively studied in \cite{AhnKooii}.

\section{the SOS model}

We have considered in the beginning that the heights take values from $1$ to
$p-1$, which is necessary for the bulk scattering weights to be finite as 
the parameter $\pi/\gamma=p$ is a positive integer. When $\pi/\gamma$
is not a rational number, there is no bounds on the heights and the
corresponding representation is known as solid-on-solid (SOS). 
The removal of the heights'
restriction certainly affects to set $\epsilon_p =1$
in the solutions (\ref{eq:udn}) of the BYBE.
Thus the solutions of BYBE in the SOS representation are
\begin{eqnarray}
U_a(u)&=&A\sin(2u+a\gamma)+\frac{B}{\sin 2u}
+\frac{\epsilon_{a-1}C}{\sin a\gamma}
\nonumber
\\
&&+\frac{D}{\sin a\gamma}
\left\{ \frac{\sin(2u+a\gamma)}{\sin 2u} -1\right\}
\nonumber
\\
D_{a}(u)&=&A\sin(2u-a\gamma)+\frac{B}{\sin 2u}
-\frac{\epsilon_{a-1} C}{\sin a\gamma}
\nonumber
\\
&&-\frac{D}{\sin a\gamma}
\left\{ \frac{\sin(2u-a\gamma)}{\sin 2u} -1\right\}
\nonumber
\\
X^{a}_{a-1,a+1}X^{a}_{a+1,a-1}&=&
\epsilon_aX^2_{13}X^2_{31}
+A^2\left\{ \sin^2 (1+\epsilon_a)\gamma -\sin^2  a\gamma\right\}
\\
&&-2AB\left\{\cos (1+\epsilon_a)\gamma
-\cos a\gamma \right\}
\nonumber
\\
&&+\epsilon_{a-1} C^2
\left\{ \frac{1}{\sin^2 \gamma} -\frac{1}{\sin^2 a\gamma} \right\}
\nonumber
\\
&& -\epsilon_{a-1}CD
\left\{\frac{1}{\cos^2 \frac{\gamma}{2}}
-\frac{1}{\cos^2 \frac{a\gamma}{2}}\right\}
\nonumber
\\
&&+D^2\left\{\frac{1}{\cos^2 \frac{(1+\epsilon_a)\gamma}{2}}
-\frac{1}{\cos^2 \frac{a\gamma}{2}} \right\}\;,
\nonumber
\label{eq:udxns}
\end{eqnarray}
redefining $C$ as $C-2D$.

Inserting above solution to the unitarity condition restricts
the coefficients in the same way as RSOS(p).
We classify the solutions into three classes:
\\
\\
Class (I) ~~~$C=0,\;A=1$
\begin{eqnarray}
U_a(u)&=&\sin(2u+a\gamma)+\frac{B}{\sin 2u}
+\frac{D}{\sin a\gamma}
\left\{ \frac{\sin(2u+a\gamma)}{\sin 2u} -1\right\}
\nonumber
\\
D_{a}(u)&=&\sin(2u-a\gamma)+\frac{B}{\sin 2u}
-\frac{D}{\sin a\gamma}
\left\{ \frac{\sin(2u-a\gamma)}{\sin 2u} -1\right\}
\\
X^{a}_{a-1,a+1}X^{a}_{a+1,a-1}&=&
\sin^2 \gamma -\sin^2  a\gamma
-2B\left(\cos \gamma -\cos  a\gamma\right)
\nonumber
\\
&&+D^2
\left( \frac{1}{\cos^2\frac{\gamma}{2}}
-\frac{1}{\cos^2 \frac{a\gamma}{2}} \right)\;.
\nonumber
\end{eqnarray}

The overall
factor $\cR_0$ is the same as that of RSOS(p),
but $\cR_1(u)$ now contains all the
information of the boundary conditions and has to satisfy
\begin{eqnarray}
&&\cR_1(u)\cR_1(-u)
\left( -\sin^2 2u -2D\cos 2u+\sin^2 \gamma-2B\cos \gamma
+\frac{D^2}{\cos^2\frac{\gamma}{2}}\right.
\nonumber
\\
&&\hspace{2.5cm}\left.-\frac{2BD\cos 2u}{\sin^2 2u}
-\frac{B^2+D^2}{\sin^2 2u} \right)=1
\\
&&\cR_1(u)=\cR_1(\gamma-u)
\nonumber
\end{eqnarray}
whose minimal solution is
\[
\cR_1(u)=\frac{\sigma(x_1,u)\sigma(x_2,u)\sigma(x_3,u)\sigma(x_4,u)}
{2\sigma(0,u)\sigma(\pi/2,u)} \;,
\]
where $x_1$-$x_4$ are related to $B,D$ via
\begin{eqnarray*}
\sum^4_{i=1} \cos 2x_i &=&-2D\\
\sum^4_{i>j=1} \cos 2x_i \cos 2x_j &=&
-2+\sin^2 \gamma -2B\cos \gamma+\frac{D^2}{\cos^2\frac{\gamma}{2}}\\
\sum^4_{i>j>k=1} \cos 2x_i \cos 2x_j \cos 2x_k &=&2(1+B)D \\
\cos 2x_1 \cos 2x_2 \cos 2x_3 \cos 2x_4 &=&
(\cos \gamma+B )^2+D^2\tan^2 \frac{\gamma}{2} \;.
\end{eqnarray*}
\\
\\
Class (II) ~~~$A=0,\; C=1$
\begin{eqnarray}
U_a(u)&=&\frac{B}{\sin 2u}
+\frac{\epsilon_{a-1}}{\sin a\gamma}
+\frac{D}{\sin a\gamma}
\left\{ \frac{\sin(2u+a\gamma)}{\sin 2u} -1\right\}
\nonumber
\\
D_{a}(u)&=&\frac{B}{\sin 2u}
-\frac{\epsilon_{a-1}}{\sin a\gamma}
-\frac{D}{\sin a\gamma}
\left\{ \frac{\sin(2u-a\gamma)}{\sin 2u} -1\right\}
\\
X^{a}_{a-1,a+1}X^{a}_{a+1,a-1}&=&
\frac{\left\{D(1-\cos\gamma)-1\right\}^2}{\sin^2 \gamma}
-\frac{\left\{D(1-\cos a\gamma)-\epsilon_{a-1}\right\}^2}{\sin^2 a\gamma} \;.
\nonumber
\end{eqnarray}

Now
\begin{eqnarray*}
\cR_1(u)\cR_1(-u)
\left[\frac{\{D(1-\cos\gamma)-1\}^2}{\sin^2 \gamma}+D^2
-\frac{2BD\cos 2u}{\sin^2 2u}-\frac{B^2+D^2}{\sin^2 2u}
\right] =1
\end{eqnarray*}
whose minimal solution is
\[
\cR_1(u)=\frac{\sin\gamma}{\sqrt{\{D(1-\cos\gamma)-1\}^2+D^2\sin^2\gamma}}
\;\;
\frac{\sigma(x_1,u)\sigma(x_2,u)}
{\sigma(0,u)\sigma(\pi/2,u)} \;,
\]
where $x_1,x_2$ are related to $B,D$ via
\begin{eqnarray*}
\cos x_1 \cos x_2 &=&\frac{(B+D)\sin\gamma}{2\sqrt{\{D(1-\cos\gamma)-1\}^2
+D^2\sin^2\gamma}}
\\
\cos^2 x_1 +\cos^2 x_2 &=&1+\frac{BD\sin\gamma}{\{D(1-\cos\gamma)-1\}^2
+D^2\sin^2\gamma}\;.
\end{eqnarray*}
\\
\\
Class (III) ~~~$A=C=0,\;D=1$
\begin{eqnarray}
U_a(u)&=&\frac{B}{\sin 2u}
+\frac{1}{\sin a\gamma}
\left\{ \frac{\sin(2u+a\gamma)}{\sin 2u} -1\right\}
\nonumber
\\
D_{a}(u)&=&\frac{B}{\sin 2u}
-\frac{1}{\sin a\gamma}
\left\{ \frac{\sin(2u-a\gamma)}{\sin 2u} -1\right\}
\\
X^{a}_{a-1,a+1}X^{a}_{a+1,a-1}&=&
\frac{1}{\cos^2\frac{\gamma}{2}}
-\frac{1}{\cos^2 \frac{a\gamma}{2}} \;.
\nonumber
\end{eqnarray}

Now
\begin{eqnarray*}
\cR_1(u)\cR_1(-u)
\left(\frac{1}{\cos^2 \frac{\gamma}{2}}
-\frac{2B \cos 2u}{\sin^2 2u}
-\frac{1+B^2}{\sin^2 2u}
\right)
=1
\end{eqnarray*}
with minimal solution
\[
\cR_1(u)=\cos \frac{\gamma}{2}\;\;
\frac{\sigma(x_1,u)\sigma(x_2,u)}{\sigma(0,u)\sigma(\pi/2,u)}
\;, \]
where
\begin{eqnarray*}
\cos^2 x_1 +\cos^2 x_2 &=&1+B\cos^2 \frac{\gamma}{2}
\\
\cos x_1 \cos x_2 &=& \frac{1}{2}(1+B)\cos \frac{\gamma}{2} \;.
\end{eqnarray*}

Note that the number of free parameters in the boundary reflection
matrices of both vertex (the sine-Gordon model)
and SOS Class (I),(II) representations are the same;
two for non-diagonal and one for diagonal\cite{pea,AhnKoo}.
This strongly suggests that there can exist a well-defined transformation 
between the two models even with a boundary.

\section{Hard Hexagon Model}

The paticle spectrum of the HHM consists of a triplet of fundamental kink 
states $K_{01},K_{10}$ and $K_{00}$\cite{Lass}.
The bulk $S$-matrix is given by\cite{Koub,Elle}
\begin{equation}
S^{ab}_{dc} (\theta)
={\cal U}(\theta) \left( \frac{\rho_a \rho_c}
{\rho_d \rho_b} \right)^{-\frac{ \theta}{2\pi i}}
W^{ab}_{dc} (u)
\end{equation}
with Boltzmann weights
\begin{equation}
W^{ab}_{dc}(u)=
\left(\frac{\rho_a \rho_c}{\rho_d \rho_b}\right)^{1/2}
\frac{\sin u}{\sin (\mu -u)}\delta_{bd}
+\delta_{ac} \;,
\end{equation}
where
\begin{eqnarray*}
&&
\rho_0 = 2\cos{\mu},\; \rho_1 = 1
\\
&&
\mu =\frac{\pi}{5},\; u = \frac{9i}{5} \theta
\\
&&
{\cal U}(\theta) = {\cal U}_0(\theta)\; {\cal U}_1(u)
\\
&&
{\cal U}_0(\theta) =
-\frac{\sinh \theta -i\sin \frac{\pi}{9}}
{\sinh \theta + i\sin \frac{\pi}{9}}
\;\frac{\sinh \theta + i\sin \frac{2\pi}{9}}{\sinh \theta - i\sin 
\frac{2\pi}{9}}
\\
&&
{\cal U}_1(u) = \frac{\sin (\mu - u) }{\sin (\mu + u)}
\;\frac{\sin (2\mu + u)}{\sin (2\mu -u)} \;.
\end{eqnarray*}
By mapping $a=2,3(1,4)$ of the RSOS(5) to $a=0(1)$ of the HHM, one can
reproduce the bulk $S$-matrix of the HHM from that of the  RSOS(5).
This means the RSOS(5) is homeomorphic to the HHM with
differences in the overall factor ${\cal U}(u)$ and
the relations between the spectral parameter $u$ and the
rapidity $\theta$.

These mean that two BYBE can be mapped to each other and the solutions
of the HHM can be obtained by that of the RSOS.
Writing the reflection amplitude of HHM as
\begin{eqnarray}
R^{a}_{bc}(\theta)&=&
{\cal R}(\theta)\left(\frac{\rho_b \rho_c}
{\rho_a \rho_a}\right)^{-\theta/2\pi i}
\nonumber
\\
&&\times\;\left[\delta_{b\neq c}X^{a}_{bc}(u)+\delta_{bc}
\left\{\delta_{b,a+1}U(u)+\delta_{b,a} V(u)+
\delta_{b,a-1}D(u)\right\}\right]\;,\label{eq:RH}
\end{eqnarray}
the non-diagonal solution is
\begin{eqnarray}
U(u)&=& A \sin (2u+3\mu) +\frac{B}{\sin 2u}
+\frac{D}{\sin 3\mu}
\left\{ \frac{\sin(2u+3\mu)}{\sin 2u} +1\right\}
\nonumber
\\
V(u)&=&A\sin (2u-3\mu) +\frac{B}{\sin 2u}
-\frac{D}{\sin 3\mu}
\left\{ \frac{\sin(2u-3\mu)}{\sin 2u} +1\right\}
\nonumber
\\
D(u)&=& A \sin (2u+\mu) +\frac{B}{\sin 2u}
+\frac{D}{\sin \mu}
\left\{\frac{\sin(2u+\mu)}{\sin 2u} +1\right\}
\\
X^{0}_{01}X^{0}_{10}&=&
A^2\left( \sin^2 \mu -\sin^2  3\mu\right)
-2AB\left(\cos \mu -\cos  3\mu\right)
\nonumber
\\
&&+D^2
\left(\frac{1}{\sin^2 \frac{\mu}{2}}
-\frac{1}{\sin^2 \frac{3\mu}{2}} \right)\;,
\nonumber
\label{eq:xnh}
\end{eqnarray}
which can be easily read from eqns.(\ref{eq:udn}),(\ref{eq:xn}).

The unitarity and crossing symmetry conditions now reduce to
\begin{eqnarray}
&&\cR(\theta)\cR(-\theta) D(u) D(-u)=1
\label{eq:uh}
\\
&&\cR(\pi i/2 -\theta)
={\cal U}_0(2\theta) \frac{\sin(2\mu +2u)}{\sin(2\mu -2u)}
\cR(\pi i/2 +\theta) \;.
\label{eq:ch}
\end{eqnarray}

It is remarkable that 
for the HHM  unitarity condition can not further reduce the arbitrary 
coefficients $A,B$ and $D$.
Among these the $B$ and $D$ terms are `decorated' solutions 
which can be constructed from a fundamental solution $R^{a}_{bc}$ 
by 
\begin{equation}
{\bf R}^e_{fd}(u)_{bf;cd}
=\sum_a S^{fe}_{ba}(u-u_1)S^{ae}_{cd}(u+u_1)R^a_{bc}(u).
\label{eq:dec}
\end{equation}
It is easy to check this satisfies the BYBE if
$S^{fe}_{ba}(u)$ is the solution of the
bulk YBE with arbitrary $u_1$.
Using the trivial solution $R^a_{bc}\propto\delta_{bc}$, one can check that
$B$ and $D$ terms can be obtained in this way. 
We will set therefore $B=D=0,\;A=1$ from now on.

The overall factor $\cR(\theta)$ is can be determined
from eqns.(\ref{eq:uh}),(\ref{eq:ch}).
Let
\begin{equation}
\cR(\theta)= \cR_0(\theta) \cR_1 (u)
\end{equation}
such that
\begin{eqnarray}
&&
\cR_0(\theta) \cR_0(-\theta) =1,
\cR_0(\pi i/2 -\theta) =
{\cal U}_0(2\theta) \cR_0(\pi i/2 +\theta)
\nonumber
\\
&&\cR_1(u) \cR_1(-u)\left(-\sin^2 2u +\sin^2 \mu \right)=1
\\
&&
\cR_1(\mu/2-\pi -u)
=\frac{\sin(2\mu +2u)}{\sin(2\mu -2u)}
\cR_1(\mu/2 -\pi+u)\;,
\nonumber
\end{eqnarray}
then the minimal solutions are
\begin{eqnarray*}
\cR_0(\theta) & = &
\frac{\sinh(\frac{\theta}{2}+\frac{\pi i}{4})}
{\sinh(\frac{\theta}{2}-\frac{\pi i}{4})}
\frac{\sinh(\frac{\theta}{2}-\frac{\pi i}{36})}
{\sinh(\frac{\theta}{2}+\frac{\pi i}{36})}
\frac{\sinh(\frac{\theta}{2}+\frac{5\pi i}{18})}
{\sinh(\frac{\theta}{2}-\frac{5\pi i}{18})}
\frac{\sinh(\frac{\theta}{2}+\frac{\pi i}{18})}
{\sinh(\frac{\theta}{2}-\frac{\pi i}{18})}
\frac{\sinh(\frac{\theta}{2}+\frac{7\pi i}{36})}
{\sinh(\frac{\theta}{2}-\frac{7\pi i}{36})}
\\
{\cal R}_1(u)&=&\frac{1}{\sin (\mu+2u)}\;.
\end{eqnarray*}

By generalizing the mapping of the RSOS(5) to the HHM,
we can construct some generalized HHM
whose particle spectrum consists of kinks $K_{ab}$
where $|a-b|=1$ with $a,b=0,\ldots,n-1$ and $K_{00}$.
Referring these as HHM(n),
the bulk $S$-matrix of the HHM(n) can be obtained from
that of the RSOS(2n+1) by folding the heights as
\begin{equation}
a,\;2n+1-a \;\;\rightarrow \;\;n-a \;\;; 1\leq a\leq n \;.
\label{eq:folding}
\end{equation}
The above one is well-defined without ambiguity,
due to the symmetry of the $S$-matrix.
The integrability conditions such as the YBE and BYBE
are transformed accordingly maintaing the structure.
These mean that we can write
the reflection amplitude of HHM(n) as
\begin{eqnarray}
R^{a}_{bc}(\theta)&=&
{\cal R}(\theta)\left(\frac{\rho_b \rho_c}
{\rho_a \rho_a}\right)^{-\theta/2\pi i}
\nonumber
\\
&&\times\;\left[\delta_{b\neq c}X^{a}_{bc}(u)+\delta_{bc}
\left\{\delta_{b,a+1}U_a(u)+\delta_{b,a} V(u)+
\delta_{b,a-1}D_a(u)\right\}\right]\;,\label{eq:RHn}
\end{eqnarray}
where $\rho_a$ denotes the $q$-number,
\[
\rho_a=\frac{\sin \bar{a}\lambda}{\sin\lambda}\qquad 
\bar{a}=n-a \qquad
\lambda=\frac{\pi}{2n+1} \;.
\]
Then the BYBE's solution for HHM(n) are
\begin{eqnarray}
U_a(u)&=&(-1)^{\bar{a}+1}A\sin(2u-\bar{a}\lambda)+\frac{B}{\sin 2u}
\nonumber
\\
&&-\frac{D}{\sin \bar{a}\lambda}
\left\{ \frac{\sin(2u-\bar{a}\lambda)}{\sin 2u} -(-1)^{\bar{a}}\right\}
\nonumber
\\
D_{a}(u)&=&(-1)^{\bar{a}+1}A\sin(2u+\bar{a}\lambda)+\frac{B}{\sin 2u}
\nonumber
\\
&&+\frac{D}{\sin \bar{a}\lambda}
\left\{ \frac{\sin(2u+\bar{a}\lambda)}{\sin 2u} -(-1)^{\bar{a}}\right\}
\\
V(u)&=&(-1)^{n+1}A\sin(2u+n\lambda)+\frac{B}{\sin 2u}
\nonumber
\\
&&+\frac{D}{\sin n\lambda}
\left\{\frac{\sin(2u+n\lambda)}{\sin 2u}-(-1)^n\right\}
\nonumber
\\
X^{a}_{a-1,a+1}X^{a}_{a+1,a-1}&=&
A^2\left\{ \sin^2 \lambda -\sin^2  \bar{a}\lambda\right\}
-2AB\left\{\cos \lambda -(-1)^{\bar{a}+1}\cos \bar{a}\lambda \right\}
\nonumber
\\
&&+2D^2
\left\{ \frac{1}{\sin^2 \lambda}-\frac{1}{\sin^2 \bar{a}\lambda}\right\}
+2D^2
\left\{\frac{\cos \lambda}{\sin^2 \lambda}
-(-1)^{\bar{a}+1}\frac{\cos \bar{a}\lambda}{\sin^2\bar{a}\lambda} \right\}
\;,
\nonumber
\end{eqnarray}
which can be read from eqns.(\ref{eq:udn}),(\ref{eq:xn}) with odd $a$.

\section{Conclusion}

In this paper we derived the most general nondiagonal reflection matrices 
of the RSOS/SOS models and the hard hexagon model
using the boundary Yang-Baxter equations. 
We find new one-parameter family of the reflection matrices 
for the RSOS model which generalizes the previous result 
in \cite{AhnKoo} where there is no free parameter.
This free parameter can be used to control the flow between
the fixed and free boundary conditions.
Since the bulk RSOS theory describes the perturbed conformal theories
by the least relevent operator, the boundary conditions can show
how the conformal boundary conditions can change under renormalization group
flows.

It is still open problem to show the vertex-face correspondence under the
presence of the boundary between the sine-Gordon and the SOS theories
while our three classes of the SOS reflection matrices 
may be useful for this purpose.

For the  hard-hexagon model which can be mapped to RSOS(5) model by folding
four RSOS heights into two,
the solutions can be obtained similarly with a main difference
in the boundary unitarity conditions.
Due to this, the reflection matrices can have two free parameters.
We show that these extra terms can be
identified with the `decorated' solutions.
This means the general solution space of the BYBE is spanned by
each fundamental solutions and decorated ones.

Considering that the HHM is related to the perturbed conformal theory
by the most relevent operator, it will be interesting to consider
how the two different perturbations can make difference in the 
boundary interactions.

\section*{Acknowledgement}

CA is supported in part by KOSEF 961-0201-006-2,  
BSRI 97-2427 and  by a grant from KOSEF through CTP/SNU. 
CKY is supported by a postdoctoral grant through Korea 
Research Foundation.

\end{document}